\documentstyle{article}

\begin{document}

\begin{center}
{\huge\bf  A topological approach to renormalization and its geometrical, dimensional consequences}
\end{center}

\vspace{1cm}
\begin{center}
{\large\bf
F.GHABOUSSI}\\
\end{center}

\begin{center}
\begin{minipage}{8cm}
Department of Physics, University of Konstanz\\
P.O. Box 5560, D 78434 Konstanz, Germany\\
E-mail: farhad.ghaboussi@uni-konstanz.de
\end{minipage}
\end{center}

\vspace{1cm}

\begin{center}
{\large{\bf Abstract}}
\end{center}
The necessity of renormalization arises from the infinite integrals which are caused by the discrepancy between the orders of differential and integral operators in the four dimensional QFTs. Therefore in view of the fact that finiteness and invariant properties of operators are their topological aspects, essential renormalization tools to extract finite invariant values from those infinities which are comparable with the experimental results, e. g. regularization, perturbation and radiative corrections follow some topological standards. In the second part we consider dimensional and geometrical consequences of topological approach to renormalization for the geometrical structure and degrees of freedom of renormalized theory. We show that regularization and renormalization of QED are performed only by certain restrictive dimensional conditions on QED fields. Further it is shown that in accord with our previous topological approach to renormalization of QED the geometrical evaluation of applied dimensional renormalization conditions and the appearance of anomalies refer to a reduction of number of degrees of freedom according to the reduced symmetry of QED. A conclusion concerning a comparison of our results with holographic principle models is also included.
\begin{center}
\begin{minipage}{12cm}

\end{minipage}
\end{center}

\newpage
The necessity of renormalization arises from the discrepancy between the orders of differential operators/ propagators which are up to two/ three and the order of integral operators in the four dimensional (4D) theories which is four. Insofar the renormalization can be considered as a method of adjustment between these different orders. In view of the fact that invariant properties of operators on suitable compact manifolds are described by their analytical/ topological indices through the dimensions of related co/homology groups therefore any invariant treatment of operators properties should be considered as a topological matter \cite{erkl}. Thus also the values of propagator integrals/ Green's functions as the inverse of differential operators in the renormalized theory should be topologically invariant in order to be compared with the globally invariant experimental values.

To justify the topological approach to renormalization of QED let us note that the Hodge-de Rham theory of differential topology is a geometric generalization of equations of classical electrodynamics (CED) \cite{w2}. Therefore the renormalization of QED becomes related with these topological aspects of CED and QED. Note that also the introduced abstract vector method in the standard renormalization calculations resulting in the Weinberg-Dyson convergence theorem \cite{wein} reminds the relevance of abstract topological methods in the renormalization. Note also that beyond the enormous relevance of topological methods in physics which appeared in the last decades, Hermann Weyl already showed one century ago that even the most empirical aspects of electrodynmaics such as the Kirchhoff laws follow purely topological laws \cite{w1}. Thus Kirchhoff followed also early topological considerations to derive them.

Furthermore note that the restriction of our topological approach to the abelian case of QED, i.e. the absence a topological approach for renromalization of non-abelian cases is due to the impossibility of a generalization of Hodge-de Rham theory of differential topology to the non-abelian cases in view of the absence of differential topological invariants, i. e. the absence of harmonic forms in these cases \cite{y}.

Nevertheless our topological approach is an attempt to make renormalization more intelligible. In other words the aim of this work is to understand why renormalization is an admissible method to extract finite values from infinite results of QED interactions and how renormalization can be understood by geometric physical consideration including topology. Thus the direct relevance of topology to understand renormalization arises from the fact that in both standard renormalization of QED and in our topological approach to renormalization one considers all physical quantities of the same dimension in geometric units, e. g. the $\displaystyle{\frac{1}{L}}$ dimensional momentum component of electron $p_\mu$, its mass $M \propto p_\mu$ and the gauge potential $e A_\mu$ as qualitatively equivalent quantities contributing to the renormalized mass of electron and to the self energies of electron and photon. For example the extraction of mass correction of electron from the radiative corrections shows that the photon represented by $A_\mu, \ F_{\mu \nu}$ participates directly to the corrected value of electron mass. Then the mass operator of interacting electron is given as the sum of two terms $M(x, x') = m_0 \delta (x-x') + ie^2 \gamma_\mu G(x, x') \gamma_\mu D_+ (x-x')$ where $G(x, x')$ is the Green's function of the Dirac equation in the external field, and $D_+ (x-x')$ is a photon Green's function \cite{sch}. Insofar both the $\displaystyle{\frac{1}{L}}$ dimensional momentum of electron and the coupled photon field $p_\mu$ and $A_\mu$ in the renormalized Dirac equation participate qualitatively equivalently to the mass correction. Note that the momentum $p_\mu$ is considered as the component of symplectic potential one form $p_\mu dx^\mu$ also in the symplectic geometry \cite{ab}. Therefore in view of the equivalent treatment of differential forms of the same order $\omega^r$ also in the topology we treat physical quantities such as momentums, masses and gauge potentials as components of their differential one forms equivalently in our topological approach to renormalization.

The main consequence of these considerations is that from topological stand point due to the invariance of lagrangian all various participants of the same geometric dimension $\displaystyle{\frac{1}{L^r}}$ in renormalization relations should be considered as components of differential forms of the same order and qualitatively equivalent quantities.

It is not so hard to consider the renormalization as a topological problem if one compares and interprets the renomalization tools such as Green's functions as differential topological quantities. Then in the topology where one treats topological invariants such as the dimensions of co/homology groups of certain differential forms, the relations among participants are perfectly adopted to achieve invariants. Thus the differential forms (forms) are dimensionally invariants in view of the $\displaystyle{\frac{1}{L^r}}$ dimensionality of their $\omega_{m_1, ...m_r}, r \in \mathbf{Z}$ components and the $L^r$ dimensionality of their $dx^{m_1} \wedge ... \wedge dx^{m_r}$ basis according to $\displaystyle{\frac{1}{L^r} . L^r} = L^0$. Therefore integrals of certain differential r-forms over certain r-chains are invariant integers known as their periods or dimensions of the related co/homology groups \cite{erkl}. Nevertheless in 4D QFTs where the integrals are 4 dimensional such as $\int_0 ^\infty ... \ dp_x \wedge dp_y \wedge dp_z \wedge dp_t$ and the propagators like $\displaystyle{\frac{1}{p^2 + ...}}$ can be considered as components of some two forms $\displaystyle{\frac{1}{p^2 + ...}} dp_x \wedge dp_y \in \omega^2$ the value of integrals of two forms over 4D volumes are divergent. This is as mentioned above due to the discrepancy between the order of relevant differential operators/ propagators and integral operators in 4D QFT.

Note that the regularization of these divergencies results in $\int_0 ^\Lambda \displaystyle{[(\frac{1}{p^2 + ...})} - \displaystyle{(\frac{1}{p^2 + ... + \Lambda^2})]} dp_x \wedge dp_y \wedge dp_z \wedge dp_t \sim \int_0 ^\Lambda \omega^4$ which is logarithmic divergent in the limit $\Lambda \rightarrow \infty$.

Moreover with respect to the invariant properties of differential operators note that their main invariant property is their topological/analytical index on a suitable compact manifold \cite{erkl}. Thus it seems that the compactification of integration manifold in QFTs by regularization /cut off is related to the requirement of a compact manifold in order to define topologically well behaved differential operators/ propagators in QFTs.

We will show in the following that essential renormalization tools to extract finite results from the infinite integrals, i. e. perturbation, regularization and radiative corrections follow some topological methods in view of the fact that finiteness, invariant properties of operators are topological aspects.

To begin note that from topological point of view one may consider all $\displaystyle{\frac{1}{L}}$ dimensional physical quantities such as the Hamiltonian $H$, components of momentum of electron, photon $p_\mu, \ k_\mu$ or gauge potential components $A_\mu$ and masses $M \sim p_\mu$ as vector components of their one forms $\omega^1$, respectively in the geometric units where all constants $\hbar, C, e$ and the velocity $v_\mu$ are dimensionless. Therefore from topological stand point it is admissible and reasonable that one renormalizes $\displaystyle{\frac{1}{L}}$ mass by corrections of $\displaystyle{\frac{1}{L}}$ Hamiltonian or momentum or gauge potential terms.

In general the topological invariant aspects of any $\displaystyle{\frac{1}{L^r}}$ dimensional physical quantity can be considered as the experimentally measured invariant aspects of some tensor components of related r-form whereby we have to do with quantities up to $r= 2$, i. e. the field strengths or curvature components. Thus treating physical field strengths as differential topological two form $F = F_{\mu \nu} dx^\mu \wedge dx^\nu$ on a compact oriented manifold one can describe the experimentally measured integral form of Maxwell equations $\int_{2D, \ or \ \partial (2D)} F \propto 0, \ Q, \ or J, \ etc.$. Also the only quantum invariant which is experimentally well confirmed, i. e. the quantum of magnetic flux $\int_{2D} F \propto \hbar$ follows the same topological invariant property that integrals of suitable two forms over suitable two manifolds are as inner products $< , >$ invariant \cite{erkl}.

In the following we explain how any essential renormalization tool follows some topological method.

 As an example of application of topology in renormalization note that the radiative corrections of electromagnetic potential $A_\mu \in \omega^1$ given by $A_\mu = A^0 _\mu \oplus \Box A_\mu \oplus \Box^2 A_\mu \oplus ...$ \cite{schbook 1} follows the topological property of Hodge decomposition theorem for connection one form $\omega^1 = Harm^ 1 \oplus d \omega^0 \oplus d^\dag \omega^2$ including the decompositions of two forms and zero forms on a compact oriented manifold without boundary $\omega^2 = d \omega^1 \oplus d^\dag \omega^3 \oplus Harm^2$ and $\omega^0 = Harm^ 0 \oplus d^\dag \omega^1$. Thus after the iteration of decomposition of one form and inserting the decompositions of zero form and two form one obtains the desired relation for the mentioned radiative correction $\omega^1 = Harm^ 1 \oplus \Box \omega^1 \oplus \Box^2 \omega^1 \oplus ..., \ \Box := d d^\dag + d^\dag d$ \cite{erk2}. Here, as usual, the Lorenz gauge is assumed $d^\dag \omega^1 \equiv 0$. Thus the application of Laplace operator $\Box$ on a differential form does not change its order $\Box^n \omega^r \in \omega^r$ \cite{erkl}. Therefore the structure of radiative corrections according to \cite{schbook 1} follows from the Hodge decomposition theorem for differential forms on a compact oriented manifold without boundary whereby $A^0 _\mu dx^\mu$ represents the harmonic one form $Harm^1$. Thus also the counter term technics to complete a concrete term follows the same decomposition theorem schema for $\omega^r$ where the decomposition scheme includes complete relevant terms related to a concrete form on a compact oriented manifold without boundary.

Furthermore note that as mentioned above the necessary compactness of manifold in topological consideration of renormalization to apply the Hodge decomposition for the radiative corrections may be related to the regularization scheme of renormalization. Thus the standard regularization by cut off compactify the domain of integration. Insofar regularization follows the topologically necessary compactification of the underlying manifold in order to apply topological methods in renormalization. Then most of essential topological methods, e. g. Hodge decomposition apply only to compact oriented manifolds without boundary. Thus the regularization of QFTs is also the preparation step to renormalization in the same manner that its equivalent compactification of manifold is the preparation of manifold for the application of topological methods.

Also the general theory of perturbation $H = H_0 + H_1 + ... + H_n; \ \{H, H_i \sim \displaystyle{\frac{1}{L}} \} \in \omega^1$ follows the topological method of Hodge decomposition $\omega^1 = d^\dag d \omega^1 \oplus (d^\dag d)^2 \omega^1 \oplus ...$ related with the above mentioned iteration of Hodge decomposition theorem $\omega^1 = Harm^ 1 \oplus \Box \omega^1 \oplus \Box^2 \omega^1 \oplus ...$ under the suitable condition $d^\dag \omega^1 = 0$ on a compact oriented manifold without boundary \cite{erk2}. Here $H_0 \in Harm^1, \ H_i \in \Box^i \omega^1$. Thus also the subtraction method in regularization technics can be considered topologically as subtraction of two equivalent components $\displaystyle{(\frac{1}{p^2 + ...})} - \displaystyle{(\frac{1}{p^2 + ...})}$ of some two forms by $\omega^2 = d^\dag d \omega^2 \oplus (d^\dag d)^2 \omega^2$ following the Hodge decomposition of two forms $\omega^2 = d\omega^1 \oplus Harm^2$, where the Hodge decomposition of one forms $\omega^1 = d^\dag \omega^2 \oplus ...$ is included.

In other words in view of the topological equivalence of all terms in a Hodge decompositions, e. g. $H_i \in \omega^1; \ i = 1, ...n$ or $\Box^i A_\mu, \ \in \omega^1$ the degeneracy which arise from each of these terms in any order can be compensated by degeneracy arising from the other term(s) in further order and finite results arising from one term can be completed by finite results from other(s) as required in renormalization. Accordingly the topological equivalence of terms in iterated Hodge decomposition explains the correctness of compensation of divergent terms in different orders by each other. Then one may consider the power $i$ in the Hodge decomposition, e. g. in $H_i \in \Box^i \omega^1$ or in $\Box^i A_\mu$ as the order of perturbation.

It is worth mentioning that also the usual gauge transformation $A_\mu = A^0 _\mu + d \lambda$ follows the Hodge decomposition theorem of topology for $\omega^1$ in the absence of $d^\dag \omega^2$ which recalls the absence of the matter current one form $J = d^\dag \omega^2, J \in \omega^1$ in the homogenous Maxwell equations. Whereas the continuity of vector current $J_\mu$ becomes a differential topological identity according to $d^{\dag ^2} \equiv 0$ \cite{erkl}. Further note that also Lorenz gauge condition is a special case of the general transversality condition $d^\dag \omega^1 = 0$ which applies on any structurally stable or topologically stable dynamical system \cite{ab}. Insofar also these aspects of physics including classical and QED follow topological standards on the mentioned compact manifolds.

Furthermore the structure of radiative correction according to Feynamn diagrams follows also the topological measure of the Euler characteristic of diagrams. Thus the second order radiative correction to replace the vertex by a triangle diagram \cite{schbook 1} is topologically admissible in view of the fact that both vertex and triangle possess the Euler characteristics $+1$. Then for a vetex alone the Euler characteristic $\chi (v) = 1_v = 1$ and for a triangle $\chi (triangle) = +3_v - 3_e +1_f = 1$ \cite{erkl} with $v$, $e$ and $f$ for the vertex, edges and faces. Because from topological stand point you do not need to distinguish between the electron $p_\mu$ or photon $k_\mu$ edges according to their above discussed differential topological equivalence as one form components. Thus also the Euler characteristic of the self energy diagram of electron \cite{schbook 1} is $2_v - 2_e + 1_f = 1$.

It is worth mentioning that all diagrams with Euler characteristic $1$, e. g. the self energy, vacuum polarisation and proper vertex part are logarithmically divergent, i. e. divergent of order one for $\Lambda \rightarrow \infty$ \cite{n}. In other words the order of divergence of these diagrams equals their Euler characteristics. Nevertheless if one assume a regularization by compactifying the integration manifold of these interactions by $\Lambda < \infty$ the interaction integrals become convergent. Topologically this would mean that in order to renormalize them their diagrams should be considered on a compact surface where two faces or closed areas exist, one inside and one outside of these diagrams which increase the Euler characteristic of these diagrams from $1$ to $2$, i. e. $+3_v - 3_e +2_f = 2$ and $+2_v - 2_e +2_f = 2$, respectively. Thus also the Euler characteristic of the usual compact model of 4D Euclidean space-time manifold $S^4$ required by Wick rotation of the renormalization theory is $2$ \cite{s}.

In the sense of general confirmation of topological approach and a further motivation regarding the use of topology in renormalization note that the forrest method in renormalization manifests the relative co/homology on a manifold with respect to its sub-manifolds \cite{rh}. Thus also the Hopf algebra approach to renormalization by trees \cite{brcokr} follows the topological methods in view of the facts that trees are special graphs and that global invariant aspects of any algebra is related to its topological aspects. For example the concepts of coproduct and antipode in the Hopf algebra approach according to the admissible cuts recalls the K\"{u}nneth formulas of standard topology.

{\large A model of topological renormalization}

The most important achievement of renormaliaztion is known to be the explanation of anomalous magnetic moment of electron and the Lamb shift related to the interaction of electron with the electromagnetic field strength and vacuum polarization according to the new term added to the Dirac equation \cite{sch}. We prove in the following that such an additional term can be explained as the Hodge decomposition theorem of the electromagnetic one form on an oriented compact manifold without boundary.

As it is mentioned above the regularization results in the compactness of the integration manifold of interactions where one is able to use the Hodge decomposition theorem after requiring its orientability and boundarylessness. Accordingly one may interpret the renormalized Dirac equation with electromagnetic potential and field strength coupling \cite{sch} as a component of the Hodge decomposition of a connection one form $Harm^1 \oplus d \omega^0 \oplus d^\dag \omega^2 - \omega^1 = 0$ multiplied by $\omega^0$ in view of the fact that $\omega^0 . \omega^r  \in \omega^r$.

\begin{equation}
( d \omega^0 \oplus \omega^1 \oplus Harm^ 1 \oplus d^\dag \omega^2) \omega^0  = 0 \ \sim \ (- \gamma [i \partial + eA] + m - \mu' . 1/2 \sigma F) \psi = 0,
\end{equation}

where the abstract $\oplus$ sum incorporates $\pm$ of related representant differential forms, $\mu' = \int ...$ is a constant value proportional to the fine structure constant and $F \in \omega^2$ is the constant magnetic field strength \cite{sch}.

Here $\psi \sim \omega^0 , \ e \gamma A \sim \omega^1 , \ i \gamma \partial \sim d , \ m \sim Harm^1, \ 1/2 \sigma F \sim \omega^2$ and hence $d^\dag \omega^2 \in \omega^1 = \int F ... = F \ . \int ...$ according to  $d^\dag F := \int \omega^2, \ \omega^1 := \int \omega^2, \ \omega^2 = d \omega^1, \ F = dA, \ A \in \omega^1$ and the constancy of $F$. The main effect is here the contribution of $d^\dag F \sim \int \omega^2 \sim \sigma F \ . \int ... \sim  \sigma F \ . \ \mu'$ of the constant magnetic field strength $F$ which is not involved in the standard Dirac equation.

The most important point in the Hodge decomposition of one form is that on the suitable compact manifold there are three terms of decomposition, i. e. the exterior differential of a zero form, the adjoint exterior differential of a two form and the harmonic one form. Altogether four terms of the same order. Exactly these four terms appear in the renormalized Dirac equation whereas the usual Dirac equation possesses only three terms. In other words the Hodge decomposition terms on the mentioned compact manifold include all necessary terms which are not all considered in differentials like the Dirac equation. Whereas they must be included by renormalization in order that the equation has finite ''compact'' values comparable with the experimental values. Thus the ''rigorous'', ''explicit'' and ''exact solutions'' for the renormalized Dirac equation are given just for the constant field strength and plane wave \cite{sch}.

Note also that the standard renormalization calculations to reproduce the experimental result of anomalous magnetic moment uses various calculative adjustments to reproduce those result \cite{sch}. Nevertheless the main achievement is as mentioned above the appearance of the additional term proportional to $F$ in the renormalized Dirac equation which can be founded by the above discussed topological necessity.

In the same manner other renormalization corrections such as the self energy corrections can be founded and understood also topologically using a. o. the Hodge decompositions of zero form for wave functions and the electromagnetic two form of Photon under suitable conditions.

In other words the renormalization/ regularization can be founded and understood topologically as the requirement of a compact oriented manifold without boundary for the space-time/ momentum space integrations where differential equations are described by Hodge decompositions why also Green's function or propagators are adjusted to dies decompositions. Because also the renormalized Green's function $ie^2 \gamma_\mu G(x, x') \gamma_\mu D_+ (x-x')$ includes the additional terms proportional to the constant electromagnetic field strength $F$ \cite{sch}.

It is interesting to mention with respect to Feynman's space-time approach to renormalization that also the Tylor expansion with which he founded the achievement of Schroedinger equation and the relevance of his approach to quantum mechanics \cite{sch} follows the mentioned Hodge decomposition theorem for zero forms if one assumes a one dimensional compact manifold where $\omega^1 = * \omega^0$ according to the Hodge duality. Thus after first iteration $\omega^0 = d^\dag \omega^1 \oplus Harm^0 \rightarrow \omega^0 = Harm^0 \oplus * d \omega^0 \oplus d^\dag d \omega^0  \sim \ f (x) = f(0) + \partial f(x) + \partial^2 f(x)$, respectively. Note that although all such expansions are performed usually up the second order nevertheless it is possible to achieve all further terms of derivatives from the further iteration of Hodge decomposition as above \cite{erk2}.

\newpage

\textbf{\large {Second part: Geometrical and dimensional consequences of topological approach to renormalization}}

In this part we will extend our topological results concerning the renormalization of QED according to the fact that dimensional conditions on tensor variables of QED such as electromagnetic vector potential or field strength tensor implies geometrical consequences for QED \cite{v}. We showed above that any essential renormalization methods such as regularization and perturbation, radiative corrections can be considered as topological methods by the use of compact spaces and iterated Hodge decomposition, respectively. Thus as in renormalization where the mass, momentum or energy of particles which are involved in an interaction are renormalized and corrected by adding of momentums and energies of their interacting fields and particles also in the differential topology all various quantities described as components of differential forms of same order are equivalent and participate to the results equivalently. Then from topological stand point due to the invariance of lagrangian all various participants of the same geometric dimension $\displaystyle{\frac{1}{L^r}}$ in renormalization formulas should be considered as components of differential forms of the same order and qualitatively equivalent quantities \cite{nak}.

The geometrical and topological aspects of renormalization begin with the very general integrability of differential equations of motion in classical field theory. Thus also the integrability or the existence of concrete solutions of equations of motion for renormalized quantities in QED requires certain conditions on these quantities \cite{sch}. Already the integrability of Schr\"{o}dinger equation in spherical coordinates $(r, \theta, \varphi, t)$ requires as Schr\"{o}dinger shows that a condition $\rho^2 + \rho = n(n+1)$ must be fulfilled for variables related to $r$ \cite{schr}. This condition is related with Bohr's quantization condition $r_n \propto n^2$ which shows that the integrability of such a differential equations requires the quantization of its main variable. Insofar the integrability of equations of motions are usually given under some remarkable geometrical conditions. We consider here oriented compact manifolds without boundary manifolds where Hodge-de Rham theory applies \cite{nak}.

In the following we will show and prove that topological aspects of renormalization and or integrability of QED restricts its geometrical structure in view of the required compactness of its underlying manifold whereby some of result may apply also to non abelian models. Thus the topological or globally admissible degrees of freedom on the compact manifold $S^4$ as the mathematical model of space-time in QFTs are also restricted. In other words one should distinguish between the assumed number of local dimensions and variables on the manifold subjected to constraints and gauge conditions and the number of topologically admissible degrees of freedom for the compact manifold of renormalized models. In order to clarify this point note on the one hand it is known that quantized models do not retain all classical symmetries in view of anomalies. On the other hand symmetries in the classical models apply on coordinates and other variables of models and the number of these symmetries depends on the number of variables of models. Therefore if quantized model does not retain all such symmetries between certain variables or coordinates it can be expected that it does not retain also the related variables or coordinates as well and the number of degrees of freedom in the quantized model is reduced in this manner. Thus topological description of renormalization to correct the quantization in the sense of finite results supports such a reduction of degrees of freedom specially in view of restriction of number of degrees of freedom on compact manifolds as the mathematical space-time models in QFT according to their Euler characteristics (see below). This result is in accord with 't Hooft's result in quantum gravity where the observable degrees of freedom are describable as Boolean variables defined on a two-dimensional lattice, evolving with time \cite{ho}.

To begin with topological and geometrical aspects of renormalization of QED note that despite of the four dimensional formulation of differential Maxwell equations the geometrical structure of classical electrodynamics (CED) described by the integral Maxwell equations are given by two dimensional surface integral and related contour integrals and two dimensional boundary surface integrals of gauge invariant field strengths: $\sim \int \int F_{\mu \nu} \propto, \ \sim \oint F_{\mu \nu} \propto \displaystyle{\frac{d}{dt}} \int \int F_{\mu \nu}, \ \mu, \nu = 1, ..., 4$. This is due to the antisymmetric second order tensor character of electromagnetic field strength, i. e. its curvature two form character that possesses only two dimensional surface integral invariants such as the magnetic quantum flux. This two dimensional surface and its boundary integral description of Maxwell equations indicate a two dimensional geometrical structure given by the Euler characteristic of certain compact two dimensional manifolds with or without boundary which are given also by similar two dimensional surface integral and or together with an additional boundary integral, respectvely \cite{bond}. Insofar the coordinate independent or invariant integral form of CED equations manifest also a two dimensional geometrical structure.

As Schwinger proved the integrability of equation of motion for renormalized electromagnetic field strength requires its constancy $F ^{em} = F ^{em} _{renorm.}: cte.$ \cite{sch}. This means in differential topological context of Hodge-de Rham theory that the constant renormalized electromagnetic two form $dF ^{em} = 0, \ F ^{em} = F ^{em} _{\mu \nu} dx^\mu \wedge dx^\nu$ is given by $F ^{em} = d \omega^1 \oplus Harm^2$. Therefore the renormalized electromagnetic two form belongs to the second cohomology group $F ^{em} \in H^2 _{em}$ in view of the fact that it is a closed but not exact form. Further for any such renormalized $F_{\mu \nu} dx^\mu \wedge dx^\nu$ resulting from the minimum of variation of an invariant electromagnetic action zero form: $S_{action} ^{em} \in \omega^0; \ \omega^0 := d^\dag \omega^1  \oplus Harm^0, \ d^\dag \omega^0 \equiv 0$ the variation of action zero form $\delta \omega^0 = 0 \Rightarrow  d \omega^0 = 0$ results in $d d^\dag \omega^1 = 0$. Note that the last relation is fulfilled using the  Lorenz gauge $\partial_\mu A_\mu = 0$, i. e. the transversality condition for the electromagnetic one form $d^\dag \omega^1 _{em} = 0$. Therefore the result of variation of electromagnetic action is $\omega^0 _{em} \in H^0$ in view of the fact that zero forms are by definition no exact forms \cite{nak}. Hence the constancy of renormalized electromagnetic two form, i. e. the existence of electromagnetic second cohomology $F_{\mu \nu} \in H^2 _{em}$ is accompanied by the zero cohomology of action form $H^0 _{em}$ and the differential topological relation: $Harm^0 _{em} \Leftrightarrow Harm^2 _{em}$. Whereas the first cohomology group is absent for electromagnetic one form view of $\omega^1 _{em} = d^\dag \omega^2 \oplus Harm^1, \ d \omega^1 d d^\dag \omega^2 \neq 0$.

The existence of only these two cohomologies $H^m, m = 0, 2$ for underlying manifold of renormalized QED, i. e. in view of constancy of renormalized electromagnetic field strength can be considered topologically as a condition which distinguishes a two dimensional oriented compact manifold similar to $S^2$ distinguished by its two cohomology classes $H^m ( S^2), \ m = 0, 2$. The compatibility of topological structure of underlying manifold of renormalized QED with the compact $S^2$ topology can be understood beyond the compatibility of regularization prescription with the compact structure of this manifold that allows only bounded integrals also according to following topological fact that in renormalized QED one can gauge away as in any gauge model the electromagnetic gauge potential or connection one form with the result that the first cohomolgy group is absent on the manifold where QED take place $H^1 (M_{QED}) = 0$. Accordingly on this QED manifold there exists as we proved above only two cohomologies $H^m, m = 0, 2$. In other words also the physical possibility of gauge away of gauge potentials underlines the compatibility of topological structure of underlying manifold of renormalized QED $M_{QED}$ with that of a compact $S^2$. Note also that as we discuss in the following there are similar $\displaystyle{\frac{1}{L^2}} = 0$ dimensional Lorenz gauge conditions applied to insure the integrability of CED differential equations via Green's functions in the same manner as the Schwinger's constant filed condition insure the integrability of QED equations \cite{sch}.

In this manner the differential topology of remormalized QED restricts the topological structure of its underlying mathematical model of space-time to an oriented compact boundaryless space with only two degrees of freedom in accord with the fact that also the degrees of freedom for originally assumed $S^4$ model of space time is two \cite{nak}. This result is supported by our previous result that the cut off or regularization prescription of renormaliaztion is geometrically and topologically equivalent to a compactification of space-time domain of QED to $S^4$ with the hint that it possesses only two degrees of freedom according to its Euler characteristic as discussed in the first part. In other words the Schwinger's condition for constancy of electromagnetic field strength as the renormalization condition of QED becomes from geometrical/topological stand point equivalent to a geometrical reduction of degrees of freedom of model.

Further facts such as the exact solvability of Dirac equation for plane waves and the identity of physical quantities characterizing the plane wave field with those of constant electromagnetic field mentioned in \cite{sch} underline our topological result that exact solvable or integrable QED equations manifests two degrees of freedom in view of the two degrees of freedom of plane wave manifested by the two degrees of freedom of photon. The geometrical aspect of such a dimensional reduction by renormalization of QED can be understood also beyond the Schwinger's condition from various other dimensional conditions for renormalization of QED such as those in renormalization methods of Thomonaga and Feynman, in regularization methods of Bethe and Pauli-Villars and the dimensional Lorenz gauge condition and Ward identity. Then the fulfillment of these dimensional methods and conditions distinguish geometrically certain lower dimensional domains in the original domain of variables of theory where the mentioned dimensional conditions are fulfilled (see below).

To underline these geometrical results let us recall that in the current anomaly the current plays two different  topological roles, i. e. as the differential one form and its Hodge dual differential three form. Therefore we repeat first some simple facts about differential forms according to which the dimensional invariance of a differential form $\omega = \omega_{1, ..., r} \ dx^1 \wedge ...\wedge dx^r$ requires the $\displaystyle{\frac{1}{L^r}}$ dimensionality of its $\omega_{1, ..., r}$ component in view of the obvious $L^r$ dimensionality of its coframe base $dx^1 \wedge ...\wedge dx^r$ \cite{nak}. Hence note that the chiral current vector $J^\mu _5 = \bar{\psi} \gamma^5 \gamma^\mu \psi$ is a $\displaystyle{\frac{1}{L^3}}$ dimensional quantity according to the $\displaystyle{\frac{1}{L^{3/2}}}$ dimensionality of $\psi$ and invariance of minimally coupled action functional in QED. Therefore it should be considered in the sense of differential topology formally as the component of three form $J: = \varepsilon_{\mu \nu \lambda \gamma} J^\mu _5 dx^\nu dx^\lambda dx^\gamma$ in order that $J^\mu _5$ to be of the required $\displaystyle{\frac{1}{L^3}}$ dimension. Nevertheless the $J^\mu _5$ is also the vector component of a one form $\tilde{J} = J^\mu _5 dx^\mu$ in view of the fact that its divergence $\partial_\mu J^\mu _5$ given by the adjoint exterior derivative $d^\dag \tilde{J} \sim d^\dag \omega^1 \in \omega^0$ is a zero form dual to the well known anomaly four form component $\partial_\mu J^\mu _5 = \varepsilon_{\mu \nu \lambda \gamma} F^{\mu \nu} F^{\lambda \gamma} \in \omega^2 _{em} \wedge * \omega^2 _{em}$. Then the divergence of chiral current as component of  three form $J: = \varepsilon_{\mu \nu \lambda \gamma} J^\mu _5 dx^\nu dx^\lambda dx^\gamma \in \omega^3$, i. e. its adjoint exterior derivative would be a two form $d^\dag \omega^3 \in \omega^2$ \cite{nak}. According to non vanishing of this divergence the quantum theory QED does not retain the conservation of chiral current or the chiral symmetry of classical lagrangian in view of the this anomaly related with the renormalization contributions \cite{abc}. In other words on the one hand the renromalization of QED or the adoption of infinite QED results into finite experimental values necessitates a reduction of the classical symmetries to less quantum symmetries according to these kind of anomalies. On the other hand this anomaly is manifested topologically by the equivalence or reduction of mentioned three form character of chiral current $J \in \omega^3$ to its one form $\tilde{J} \in \omega^1$ with its vectorial components. According to the above discussed invariance of differential forms this equivalence of current three form and one form $J \cong \tilde{J}$ results in the equality of $\displaystyle{\frac{1}{L^3}} = \displaystyle{\frac{1}{L^1}}$ dimensions for related components. This result means from geometrical and dimensional stand point that $\displaystyle{\frac{1}{L^2}} = 1$. In other words the QED should possesses constant quantities of dimension $\displaystyle{\frac{1}{L^2}} = cte.$ according to renormalization and loosing to much classical symmetries and their related degrees of freedoms. Such a $\displaystyle{\frac{1}{L^2}}$ quantity is already known in Schwinger's renormalization method to be the above discussed constant $\displaystyle{\frac{1}{L^2}}$ dimensional electromagnetic field strength $F_{\mu \nu}$. Thus $F_{\mu \nu} \in \omega^2 _{em} := F_{\mu \nu} dx^\mu \wedge dx^\nu$. Hence the last reduction of degrees of freedom in view of the reduction of related symmetry by the anomaly is in best accordance with the above discussed reduction in view of the constancy of field strength that distinguishes a $S^2$ manifold with two degrees of freedom.

The relevance of this reduction can be understood also in view of the fact that the vanishing of above divergence of chiral current one form $ \partial_\mu J^\mu \in d^\dag \tilde{J} \sim d^\dag \omega^1$ would require the vanishing of the exterior derivative of its dual three form $d^\dag \omega^1 = * d \omega^3 = 0$. Accordingly it would require that the vanishing of electromagnetic four form $\varepsilon_{\mu \nu \lambda \gamma} F^{\mu \nu} F^{\lambda \gamma} \in (\omega^2 \wedge * \omega^2)_{em} = 0$. A non trivial solution of this condition is given by $\omega^2 _{em} \equiv * \omega^2 _{em}$ or, i. e. by $\omega^2 _{em} = \omega_{mn} dx^m \wedge dx^n, \ m, n = 1, 2$ or $\omega^2 _{em} := F_{mn} dx^m \wedge dx^n$. Then there is only one two form in two dimensions and in view of $* \omega^2 = \omega^0$ in two dimensional case, i. e. $* F_{mn} dx^m \wedge dx^n \in \omega^0$ the required vanishing of four form $(\omega^2 \wedge * \omega^2)_{em} = 0$ is given in view of the Hodge duality: $* \omega^2 _{em} = \omega^0 _{em}$ in two dimensions. In other words the success of reormalization of QED avoiding the chiral anomaly would require the reduction of the domain of QED as above to a two dimensional $S^2$.

\

To understood the geometrical character of renormalization conditions that underline the mentioned reduction note that as we mentioned above different aspects of renormalization are performed by application of various dimensional conditions on QED which are not yet take into account geometrically and topologically. Note that there are also dimensional conditions in the renormalization of non abelian gauge theory of electroweak interactions similar to that in the renormalization of abelian QED, i. e. in both cases of the order $\displaystyle{\frac{1}{L^2}} = 0$. Thus there are also similar dimensional conditions such as Lorenz gauge performed already in CED to ensure the existence of solutions or integrability of CED differential equations by Green's functions. This fact underlines the relation between the integrability of CED/QED differential equations and $\displaystyle{\frac{1}{L^2}} = 0$ dimensional conditions such as Lorenz gauge condition for CED/QED and the above mentioned $\displaystyle{\frac{1}{L^2}} = cte.$ dimensional Schwinger's slowly varying/constant filed strength condition for QED. In other words in both cases the integrability of differential equations are achieved by similar $\displaystyle{\frac{1}{L^2}}$ dimensional conditions. Thus the difference between constant value in case of QED condition and zero value in case of CED condition can be understood in view of the proportionality of QED measures to Planck's constant $\hbar$ and the classical limes $\hbar \rightarrow 0$ for CED measures.

We show in the following that the geometrical and topological consideration of all these dimensional renormalization conditions refers to a reduced number of degrees of freedom in the renormalized theory. For example the Pauli-Villars regularization condition $\sum_i c_i M_i ^2 = 0$ \cite{pv} for auxiliary masses $M_i$ and constant $c_i$ is a $\displaystyle{\frac{1}{L^2}} = 0$ dimensional condition in view of $\displaystyle{\frac{1}{L}}$ dimensionality of mass. Accordingly it is a dimensional condition $\displaystyle{\frac{1}{L^2}} = 0$ that distinguishes a rest two dimensional geometry $\sim \displaystyle{\frac{1}{L^2}} \sim L^2$ within the original 4D geometry where the regularization and renormalization of QED takes place. Thus also the Tomonaga method to describe a relativistically invariant QFT by four dimensional delta functions concludes that it implies an invariant $\displaystyle{\frac{1}{L^2}} = cte.$ dimensional measure $\displaystyle{\frac{dk_x dk_y dk_z}{k}}$ \cite{tom}. Therefore also this relativistic renormalization method of QED implies and distinguishes a constant two dimensional geometry $\displaystyle{\frac{1}{L^2}} = cte. \sim L^2 = cte.$ within the 4D geometry where renormalization can be applied. In the same manner also Feynman renormalization method for radiative corrections is based on the dimensional condition that requires some quadratic values of radiative corrected momentum of electrons must be neglected $q^2 = 0$ \cite{fe}. This dimensional condition is again of order $\displaystyle{\frac{1}{L^2}} = 0$ in view of the $\displaystyle{\frac{1}{L}}$ dimensionality of momentum in QFT. Accordingly also this dimensional condition distinguishes a rest $\displaystyle{\frac{1}{L^2}} \sim L^2$ two dimensional geometry remains from the original 4D geometry where the renormalization applies. Moreover also the Ward identity in renormalized QED ${\displaystyle k_{\mu }M^{\mu}=0}$ with $k_{\mu}$ the momentum of external photon and $M^{\mu}$ the correlation functions in momentum space of electron is a dimensional condition of the same $\displaystyle{\frac{1}{L^2}} = 0$ order \cite{war} that also distinguishes a rest two dimensional geometry within the original 4D geometry $\displaystyle{\frac{1}{L^2}} \ \sim L^2 \neq 0$ for renormalization of QED. Thus also Bethe's method to calculate the Lamb shift of energy levels uses a constant value for regularization of energy $K \approx mc^2$ or $m \sim \displaystyle{\frac{1}{L}} = cte$ where the mass appear in the detailed calculation of regularized energy quadratically in the square momentum term $|P_{nm}|^2$ \cite{b}. Therefore also this renormalization result is based on some dimensional constant of the same order $m^2 \sim \displaystyle{\frac{1}{L^2}} = cte.$ that distinguishes a similar $L^2 = cte.$ geometry within the original 4D geometry where the renormalization works.

Note that there are similar dimensional conditions in QED that reduce obviously the number of degrees of freedom of electromagnetic field, e. g. the Coulomb and or Lorenz gauge conditions as the electromagnetic form of transversality condition. Thus in both cases the dimensional gauge condition is given by divergencelessness of electromagnetic potential vector as mentioned above for Lorenz gauge condition $\partial_\mu A_\mu = 0 \sim \displaystyle{\frac{1}{L^2}} = 0$ in view of $\displaystyle{\frac{1}{L}}$ dimensionality of potential vector as the vector component of differential one form. Accordingly the degrees of freedom of such a gauged vector is reduced dynamically in order to achieve a transversal propagation of field also manifested by the polarization freedom of photon whereby also this dimensional conditions distinguishes a rest two dimensional geometry within the original 4D geometry in view of the rest $\displaystyle{\frac{1}{L^2}} \sim L^2 \neq 0$ structure.

A possible way to consider the two dimensional geometry distinguished by all these renormalization conditions: $L^2 = 0, \ or \ cte.$ is to consider it in 4D local coordinates as $L^2 \cong \sum_{\mu = 1} ^4 a_\mu x_\mu ^2 = 0, \ a_\mu \in \mathbf{Z}$. It can be rewritten as $\sum_{i = 1} ^3 x_i ^2 = 1$ after its normalization. The resulting relation is the equation of $S^2$ as the above mentioned distinguished two dimensional geometry with required constant curvature and area $\displaystyle{\frac{1}{L^2}} = cte. \sim L^2 = cte.$, respectively.

Insofar all discussed dimensional conditions for regularization and renormalization are consistent and distinguish some two dimensional geometry according to $\displaystyle{\frac{1}{L^2}} \sim L^2 = 0 \ or \ cte.$ within the original 4D geometry. Their consistency with each other and with the above mentioned topological reduction of $S^4, \ \chi (S^4) = 2$ to $S^2, \ \chi (S^2) = 2$ underlines the required reduction by renormalization of QED where $\chi (...)$ is the Euler characteristic of manifold.

It is worth mentioning that dimensional conditions of same order $\displaystyle{\frac{1}{L^2}} \sim L^2 = 0$ are also used in the renormalization of non abelian SU(1 x U(1) electroweak gauge model \cite{tv}.

Concluding remarks: we conclude in view of the fact that all these results about renormalized QED can be considered as quantum information which indicate a two dimensional compact geometry with two degrees of freedom comparable with a compact boundary of spatially 3D space. Therefore these renormalized QED results should be compatible with a two dimensional holographic structure of quantized information of universe discussed by holographic principle models \cite{mod}. Thus the main ingredients of holographic principle, the entropy of a black hole $S = 4 \pi M^2 + C (cte.)$ \cite{ho} is a dimensional relation $L^2 = cte.$ which relates the entropy as a number with the square of gravitational mass $\sim L$. At any rate such a dimensional relation for entropy distinguishes a two dimensional area as a constant recalling our above discussed results concerning the same dimensional aspect $\displaystyle{\frac{1}{L^2}} \sim L^2 = cte.$ where the $\displaystyle{\frac{1}{L^2}} = cte.$ was considered above as the constant curvature of some compact two dimensional manifold with constant area $L^2 = cte.$.


\begin{thebibliography}{100}
\bibitem{schbook 1}
See Dyson's works, specially "The S matrix in quantum electrodynamics" in \cite{sch}. Note that Dyson used $\Box^2$ for the d' Alambertian. Whereas we use as in the modern standard the $\Box$ for d'Alambertian which is also the Laplacian on a flat manifold of Lorentzian signature.
\bibitem{erk2}
With respect to the multiplicity of $\omega^r$ in the iteration of Hodge decomposition note that $ n . \omega^r \in \omega^r, \ n \in \mathbf{Z}$ \cite{erkl}. Therefore one may have several $\omega^r$s under operations of several powers of $d^\dag d$, etc.
\bibitem{s}
Note that if one considers the Euler characteristic on $S^4$ as the number of degrees of freedom on $S^4$ in view of the fact that Euler characteristic is equal to the index of differential operator such as the Laplacian on $S^4$ counting its independent solutions on $S^4$ \cite{erkl}. Further if one describes the propagators integrals on $S^4$ using its two degrees of freedom then these 2D integrals become convergent due to the fact that in this case the order of differential and integral operators agree with each other.
\bibitem{v}
Note that since the number of component or dimensions of tensors on a manifold depend on the assumed geometry and number of dimension of manifold therefore dimensional conditions on components of tensor variables of a physical model implies geometrical consequences for the model and its underlying manifold.
\end{thebibliography}
\end{document}